\documentclass[prb,twocolumn,amsmath,amssymb,floatfix]{revtex4}
\usepackage{graphicx}
\usepackage{bm}% bold math
\usepackage{color}

\newcommand{\be}{\begin{equation}}
\newcommand{\ee}{\end{equation}}
\newcommand{\beq}{\begin{eqnarray}}
\newcommand{\eeq}{\end{eqnarray}}

\begin{document}

\title{Universal non-equilibrium quantum dynamics in imaginary time}

\author{C. De Grandi, A. Polkovnikov, A. W. Sandvik}
\affiliation{Department of Physics, Boston University, 590 Commonwealth Ave., Boston, Massachusetts 02215, USA}

\date{December 14, 2011}

\begin{abstract}
We propose a method to study dynamical response of a quantum system by evolving it with an imaginary-time dependent Hamiltonian. The leading non-adiabatic response 
of the system driven to a quantum-critical point is universal and characterized by the same exponents in real and imaginary time. For a linear quench protocol, the 
fidelity susceptibility and the geometric tensor naturally emerge in the response functions. Beyond linear response, we extend the finite-size scaling theory of quantum 
phase transitions to non-equilibrium setups. This allows, e.g., for studies of quantum phase transitions in systems of fixed finite size by monitoring expectation
values as a function of the quench velocity. Non-equilibrium imaginary-time dynamics is also amenable to quantum Monte Carlo (QMC) simulations, with a scheme
that we introduce here and  apply to quenches of the transverse-field Ising model to quantum-critical points in one and two dimensions. The QMC method is generic 
and can be applied to a wide range of models and non-equilibrium setups.
\end{abstract}

\maketitle

\section{Introduction}

The dynamics of thermally isolated quantum systems beyond linear response has become a focus of experimental and theoretical research in 
thermalization,\cite{kinoshita, rigol_08} universal quantum critical dynamics,\cite{ap_adiabatic, zurek_05} quantum annealing,\cite{das_05} 
and many other areas.\cite{dziarmaga_10, ap_rmp} It has been realized \cite{degrandi_10} that deviations from adiabaticity in gapless systems 
and near quantum-critical points, in particular, can be characterized by scaling behavior of the fidelity susceptibility and its adiabatic generalizations. 
These susceptibilities are related to non-equal time correlations of the corresponding quench operator\cite{venuti_07, degrandi_10} evaluated either 
at the beginning or the end of the dynamical process. One can, thus, extract valuable information on the dynamical properties of quantum systems by 
analyzing their non-adiabatic response. Such response can be directly measured experimentally\cite{chin_10, chen_11} or studied numerically.
At the moment, numerical studies of real-time dynamics of interacting systems are limited to small systems, mostly in one dimension, 
however.\cite{kolodrubetz_11}

We here show that quantum dynamics can also be simulated by evolving the system in imaginary time. In particular, we demonstrate that 
the leading non-adiabatic response of a system with its Hamiltonian changing in imaginary time is very similar to that of the real-time dynamics. This 
allows us to use powerful quantum Monte Carlo (QMC) techniques to investigate the dynamical response. Another advance presented here is the extension of
the standard linear response theory, both in real and imaginary time, to non-linear driving protocols where the velocity or acceleration of the quench 
replaces the amplitude. In particular, we show that the linear response of physical observables in the case of linear quenches is characterized by the 
components of the geometric tensor.\cite{provost_80,venuti_07} This allows one to experimentally measure them, or simulate them using QMC, and study 
their singularities near quantum critical points. Previously, with QMC simulations it was only known how to compute the diagonal elements of the 
geometric tensor, i.e., the fidelity susceptibilities.\cite{albuquerque}

We show that the non-perturbative response of generic observables can be described by extending the standard finite-size scaling theory of quantum phase 
transitions to non-equilibrium protocols (e.g., by simultaneous scaling in the system size and the quench velocity, or by only changing the velocity at 
fixed system size), with exponents that we derive here. In this work we focus on imaginary time dynamics, but all universal results also apply to real-time 
protocols.

We discuss the underlying time-evolution formalism and results of adiabatic perturbation theory in Sec.~\ref{sec1}, followed by results from linear 
response theory of physical observables to the quench velocity and the emergence from it of the geometric tensor in Sec.~\ref{sec2}. Then, in Sec.~\ref{sec2a} 
we formulate the scaling theory of non-perturbative response of interacting systems to slow perturbations near quantum-critical points, extending the scaling theory 
of phase transitions to non-equilibrium setups. In Sec.~\ref{sec3} we apply the theory to the particular example of the one-dimensional transverse-field Ising model, 
where the scaling forms derived can be compared with exact results. In Sec.~\ref{sec4} we present the QMC method and numerical results obtained with it for the 
two-dimensional transverse-field Ising model. We conclude in Sec.~\ref{sec5} with a brief summary and discussion. More details of the adiabatic perturbation 
theory are given in Appendix~\ref{app1}, and properties of the dynamic susceptibilities derived are further discussed in Appendix \ref{app2}.

\section{Time evolution}
\label{sec1}

We consider the imaginary-time evolution described by a Hamiltonian $\mathcal H(\lambda)$ which implicitly depends on time through the tuning parameter 
$\lambda(\tau)$. We assume that the evolution starts at some time $\tau_0<0$ and ends at $\tau=0$, $\lambda(0)$ being the point of interest. To simplify 
the notation we set $\lambda(0)=0$. The imaginary-time propagation of the wave function in this setup is governed by the  Shr\"odinger equation 
in imaginary time:
\be
\partial_\tau \psi(\tau)=-\mathcal H(\lambda(\tau))\psi(\tau).
\label{sch_eq}
\ee
The formal solution at time $\tau$ is given by the evolution operator, $\psi(\tau)=U\psi(\tau_0)$, with $U$ given by the time-ordered exponential:
\be
U=T_\tau \exp\left[-\int_{\tau_0}^\tau d\tau' \mathcal H(\lambda(\tau'))\right].
\label{utau}
\ee

Before going into details of the dynamics, let us make some remarks: (i) In the adiabatic limit, $\dot\lambda\to 0$, the system rapidly falls into its 
instantaneous ground state, after a transient time, and then follows this state. (ii) At finite $\dot\lambda$ the system is constantly excited from the 
ground state by the evolving Hamiltonian and relaxes back due to imaginary-time propagation. The proximity to the instantaneous ground state is controlled 
by $\dot \lambda(\tau)$ near the final point $\tau\to 0$. If the velocity vanishes at the final point, $\dot\lambda(0)=0$, then the degree of nonadiabaticity 
is controlled by the acceleration $\ddot\lambda(0)$, etc. (iii) Imaginary time evolution is amenable to QMC simulations, giving access to universal aspects 
of quantum dynamics in a wide range of systems. We will outline such a generalization of standard equilibrium QMC in Sec.~\ref{sec4} and apply it to 
the transverse-field Ising model. We first discuss the analytical framework needed for analyzing both QMC and experimental results.

The easiest way to analyze the general properties of the solution of Eq.~(\ref{sch_eq}) is to go to the adiabatic (co-moving) basis. This procedure is similar 
to that in real time, though containing very important subtleties. The details of the analysis are given in Appendix \ref{app1}. Here we present only the 
final result of the first order of adiabatic perturbation theory, which contains all relevant scaling information. Denoting by $a_n(0)$ the expansion coefficient 
of the wave-function  $\psi(0)$ in the eigenstates of the final Hamiltonian we have
\be
a_n(0)\approx \int\limits^0_{-\infty} d\tau \, {\langle n|\partial_{\tau} \mathcal H|0\rangle\over \Delta_{n0}(\tau)} \exp\left[-\int^0_{\tau} d\tau'\, \Delta_{n0}(\tau')\right],
\label{eq_central}
\ee
where $\Delta_{n0}(\tau)=\mathcal E_n(\tau)-\mathcal E_0(\tau)$ is the instantaneous energy of the $n$-th level relative to the ground state and 
$\langle n|\partial_{\tau} \mathcal H|0\rangle$ is the transition matrix element between the instantaneous eigenstates.

To make further progress in analyzing the transition amplitudes (\ref{eq_central}), let us consider the very slow asymptotic limit $\dot\lambda\to 0$. To be 
specific, we assume that near $\tau=0$ the tuning parameter has the form $\lambda(\tau)\approx v |\tau^r|/r!$ (see also Ref.~\onlinecite{degrandi_10}). 
The parameter $v$, which controls the adiabaticity, plays the role of the quench amplitude ($r=0$), velocity ($r=1$), acceleration $(r=2)$ etc. It is easy to check 
that in the asymptotic limit $v\to 0$, Eq.~(\ref{eq_central}) gives
\be
\alpha_n\approx v {\langle n|\partial_\lambda|0\rangle \over (\mathcal E_n-\mathcal E_0)^r}=- 
v {\langle n|\partial_\lambda \mathcal H|0\rangle \over (\mathcal E_n-\mathcal E_0)^{r+1}},
\ee
where all matrix elements and energy levels are evaluated at $\tau=0$. From this perturbative result we can evaluate the leading non-adiabatic response of  various 
observables and define the corresponding susceptibilities.

\section{Linear Response and Geometric Tensor}
\label{sec2}

Let us represent the observables of interest as generalized forces, i.e., derivatives of $\mathcal H$ with respect to the couplings 
$\mu$; $M_\mu=-\partial_\mu \mathcal H$. By using this representation we do not lose any generality. For example, a spin-spin correlation function 
${\bf s}_i \cdot {\bf s}_j$ of some lattice model can be represented as a response with respect to an infinitesimal coupling connecting these spins. 
Then we find
\be
M_\mu = C + 2vL^d \chi_{\mu\lambda}^{(r+1)},
\label{lin_resp}
\ee 
where $C \equiv \langle \psi(0)|M_\mu|\psi(0)\rangle$ and
\be
\chi_{\mu\lambda}^{(r+1)}\!={1\over L^d}\sum_{n\neq 0}{\langle 0|\partial_\lambda \mathcal H|n\rangle\langle n|\partial_\mu \mathcal H|0\rangle+
\mu\leftrightarrow\lambda\over 2(\mathcal E_n-\mathcal E_0)^{r+1}}. 
\label{chi_g}
\ee
These susceptibilities can also be expressed through the imaginary\cite{venuti_07} and real time 
connected correlation functions,
\be
\chi_{\mu\lambda}^{(r+1)}\!\!={1\over 2L^d}\int\limits_0^\infty d\tau {\tau^r\over r!} 
\langle 0| \partial_\mu \mathcal H_\tau\partial_\lambda \mathcal H_0+\partial_\lambda \mathcal H_\tau\partial_\mu \mathcal H_0|0\rangle_c,
\label{chi_g1}
\ee
where $\partial_\lambda \mathcal H_\tau$ is the imaginary-time Heisenberg representation of the operator $\partial_\lambda \mathcal H$ evaluated at $\tau$ (and in real time 
one substitutes $\tau\to it+0^+$, as discussed in Appendix \ref{app2}). Thus, by changing the exponent $r$ of the quench protocol one can probe different moments of the 
real and imaginary time correlation functions. Let us point out that the factor $L^{-d}$ in Eqs.~(\ref{chi_g}) and (\ref{chi_g1}) is inserted for convenience for extensive 
observables, which appear as a response to global perturbations. For intensive observables this factor is not needed.

The situation is slightly different for diagonal observables like the energy,
\be
Q=\langle \mathcal H\rangle -\mathcal E_0,
\ee
 the log-fidelity,\cite{rams_11} 
\be
F=-\ln(|\langle \psi(0)|0\rangle|^2),
\ee
or the entropy entropy which in the lowest order of perturbation theory are described by quadratic rather then linear 
response:\cite{degrandi_10}
\begin{eqnarray}
Q & \approx & v^2 \chi^{(2r+1)}_{\lambda\lambda}, \nonumber \\
F & \approx & v^2 \chi^{(2r+2)}_{\lambda\lambda}.
\label{en_fid}
\end{eqnarray}
The response coefficients in Eq.~(\ref{lin_resp}) have a very interesting geometric interpretation for linear quenches ($r=1$). Then the susceptibility $\chi_{\mu\lambda}^{(2)}$ 
reduces to the symmetrized $\mu\lambda$-component of the geometric tensor,\cite{provost_80,venuti_07} which defines the Riemannian metric in the manifold of the ground 
states of the Hamiltonian $\mathcal H(\mu,\lambda)$.\cite{provost_80} The diagonal components of the geometric tensor $\chi_{\lambda\lambda}^{(2)}$ define the fidelity 
susceptibilities.\cite{gu_09} We emphasize that the metric tensor, which was originally thought to have no physical significance,\cite{provost_80} emerges here as a 
response of physical observables to the quench velocity. Thus, Eq.~(\ref{lin_resp}) opens a practical (numerical or experimental) way of analyzing the geometry of the 
ground state wave function in the parameter space and studying its universality, nature of its singularities, and its topology.

\section{Scaling Theory}
\label{sec2a}

In gapped systems all non-equal time correlation functions decay exponentially with time, implying that the susceptibilities $\chi^{(m)}_{\mu\lambda}$ converge for all $m$ 
in the thermodynamic limit. For gapless systems the situation is more complicated and the susceptibilities can diverge. To understand the nature of this divergences we will employ scaling analysis.

If the quench operator $\partial_\lambda \mathcal H$  is marginal or relevant, then its scaling dimension is,
\be
\Delta_\lambda\equiv {\rm dim}[\partial_\lambda \mathcal H]=z-{\rm dim [\lambda]}. 
\ee
For a marginal perturbation maintaining gaplessness in the vicinity of $\lambda=0$ and not affecting the dynamic exponent $z$, we have ${\rm dim}[\lambda]=0$ and $\Delta_\lambda=z$. 
Such cases include  superfluids and Fermi liquids (with $\lambda$ the interaction coupling). If $\partial_\lambda \mathcal H$ is relevant, e.g., when driving the system to a 
gapped phase at $\lambda \not=0$, then by definition ${\rm dim}[\lambda]=1/\nu$,where $\nu$ is the correlation length exponent.\cite{sachdev} Then $\Delta_\lambda=z-{1/\nu}$, 
and from Eq.~(\ref{chi_g}) we obtain 
\be
\eta^{(r+1)}_{\mu\lambda}\equiv{\rm dim}[\chi^{(r+1)}_{\mu\lambda}]=\Delta_\mu+d-1/\nu-z r. 
\ee
For $\mu=\lambda$ this expression reduces to a known result.\cite{degrandi_10,schwandt_alet_09}

If $\eta_{\mu\lambda}^{(r+1)}<0$ the susceptibility diverges with the system size, 
\be
\chi^{(r+1)}_{\mu\lambda}\sim L^{-\eta_{\mu\lambda}^{(r+1)}}, 
\ee
and the perturbative 
result~(\ref{lin_resp}) breaks down in the thermodynamic limit. To find the correct asymptotics of the observables in this case, 
we introduce the scaling dimension of the velocity; 
${\rm dim}[v]={\rm dim}[\lambda]+zr=1/\nu+zr$. Following arguments similar to Ref.~\onlinecite{degrandi_10} instead of Eqs.~(\ref{lin_resp}) and (\ref{en_fid}) we then find:
\beq
M_{\mu} & \approx & C+L^d v^{(d+\Delta_\mu)\nu\over 1+\nu z r} f_{\mu\lambda}(v L^{zr+1/\nu}) \nonumber \\
& = & C + L^{-\Delta_\mu}\tilde f_{\mu\lambda}(vL^{zr+1/\nu}).
\label{scaling}
\eeq
Here $C$ is some non-universal constant. However, unlike in Eq.~(\ref{lin_resp}), this constant does not have to be the ground state expectation value. Whether the ground 
state expectation value is included or not in $C$ determines the small velocity asymptotics of the scaling functions $f_{\mu\lambda}(x)$ and $\tilde f_{\mu\lambda}(x)$. 
In general these asymptotics can be determined from physical arguments. For $x\ll 1$ we should recover linear or quadratic response for diagonal and off-diagonal observables, 
respectively, plus possibly universal ground state contribution if it is not included in $C$. The large argument asymptotic of the scaling function can be obtained from 
other considerations. For example, for extensive operators the scaling functions should saturate at large $x$ so that $M_\mu$ is extensive. The properties of the 
susceptibilities (\ref{chi_g}) are further discussed in Appendix \ref{app2}.

Instead of the length scale equal to the system size $L$ in Eq.~(\ref{scaling}) there can be another relevant length scale, e.g., the distance $x_{12}=|{\bf x}_1-{\bf x}_2|$
between two points ${\bf x}_1$ and ${\bf x_2}$ if we are interested in correlation functions. Thus, in a translationally invariant system one expects that the non-equilibrium 
connected correlation function in a large system should scale as:
\be
\langle M_\mu({\bf x}_1) M_\mu({\bf x}_2)\rangle_c\approx {1\over x_{12}^{2\Delta_\mu}}f\left(vx_{12}^{zr+1/\nu}\right).
\ee 
Likewise we can generalize Eq.~(\ref{scaling}) to quenches which at the final time end up in the vicinity of the QCP, i.e., at $\lambda_f\neq \lambda_c$:
\be
M_{\mu}  \approx  C +L^d v^{(d+\Delta_\mu)\nu\over 1+\nu z r} f_{\mu\lambda}\left(v L^{zr+1/\nu}, {|\lambda_f-\lambda_c|\over v^{1/(z\nu r+1)}}\right).
\label{scaling1}
\ee
This scaling relation can be used for independently locating the quantum critical point by, e.g., sweeping across the phase transition in a sufficiently big system 
with different velocities (such that $vL^{z+1/\nu}\gg 1$). Then there will be a crossing point in $M_{\mu}/v^{(d+\Delta_\mu)\nu/(1+\nu z)}$ plotted versus the coupling 
$\lambda_f$ in curves corresponding to different velocities. The expression (\ref{scaling1}), which is applicable to real-time protocols as well, also suggests a 
convenient way for determining the location of the critical point experimentally, by changing the velocity for a fixed system size. The usual finite-size scaling procedure
requires changing the system size, which is not always feasible. On the other hand, changing the quench velocity would normally be quite straightforward in experiments
on, e.g., cold atoms.

The scaling relations (\ref{scaling}) generalize the standard finite-size scaling theory of quantum phase transitions (corresponding to $r=0$) to non-equilibrium 
protocols and constitute our main analytical result. Eq.~(\ref{scaling}) is valid both for real and imaginary time, and also for the expectation value $\langle M_\mu\rangle$ 
(in which case the small argument asymptotics is dictated by the requirement that $\langle M_\mu\rangle$ reduces to its equilibrium value in the adiabatic limit), as 
well for any other observable $\mathcal{O}$ with scaling dimension $\Delta_\mathcal{O}$. For example, for the particular cases of the energy and the fidelity, 
$\Delta_E=z$ and $\Delta_F=0$, respectively, and  Eq.~(\ref{scaling}) reduces to known results,\cite{degrandi_10} which were recently verified numerically for 
a particular 1D model.\cite{kolodrubetz_11}

\section{1D Tranverse-field Ising model}
\label{sec3}

To illustrate the above general scaling results we consider a linear quench in the one-dimensional (1D) transverse-field 
Ising model with Hamiltonian
\be
\mathcal H=-\sum_j \sigma_j^x-J\sum_{\langle ij\rangle}\sigma_i^z\sigma_{j}^z,
\label{is1}
\ee
where $\sigma_x$ and $\sigma_z$ are the Pauli matrices and $\langle ij\rangle$ are nearest neighbours sites. The dimensionless coupling constant $J$ drives the system through 
a critical point at $J_c=1$, with critical exponents $z=\nu=1$.  We consider the following quench protocol: $J(\tau)=1+\lambda=1+v\tau$, starting in the ground state at 
$\tau_0=-1/v$.  Using the Jordan-Wigner transformation, the model can be mapped to free fermions. The analysis of the imaginary time dynamics is straightforward and available 
in the literature for similar real-time setups.\cite{dziarmaga_05,degrandi_2009} We therefore only quote our results in Table~\ref{table1}.

\begin{table}[tc]
\begin{center}
\begin{tabular}{|c|c|c|c|}
\hline
Observable & $ E_z$ & $Q$ &  $F$\tabularnewline
\hline
 $vL^2\ll 1$ & ${1\over 16} v L^2$   & ${7\zeta(3)\over 128} v^2 L^3$ & ${1\over 6144} v^2 L^4 $ \tabularnewline
 \hline
$vL^2\gg 1$ &  $0.26\sqrt{v}L$ & $0.0265\, vL$  &  $0.0276 \sqrt{v} L$ \vspace{0.05cm}\tabularnewline
\hline
\end{tabular}
%\hfill{}
\caption{Scaling of the excess interaction energy $E_z$ with respect to the final ground state, the excess total energy $Q$, and the log-fidelity $F$ with 
the quench rate and the system size for the transverse-field Ising chain.}
\label{table1}
\end{center}
\vskip-6mm
\end{table}

\begin{figure}
\begin{center}
\includegraphics[width=8.4cm, clip]{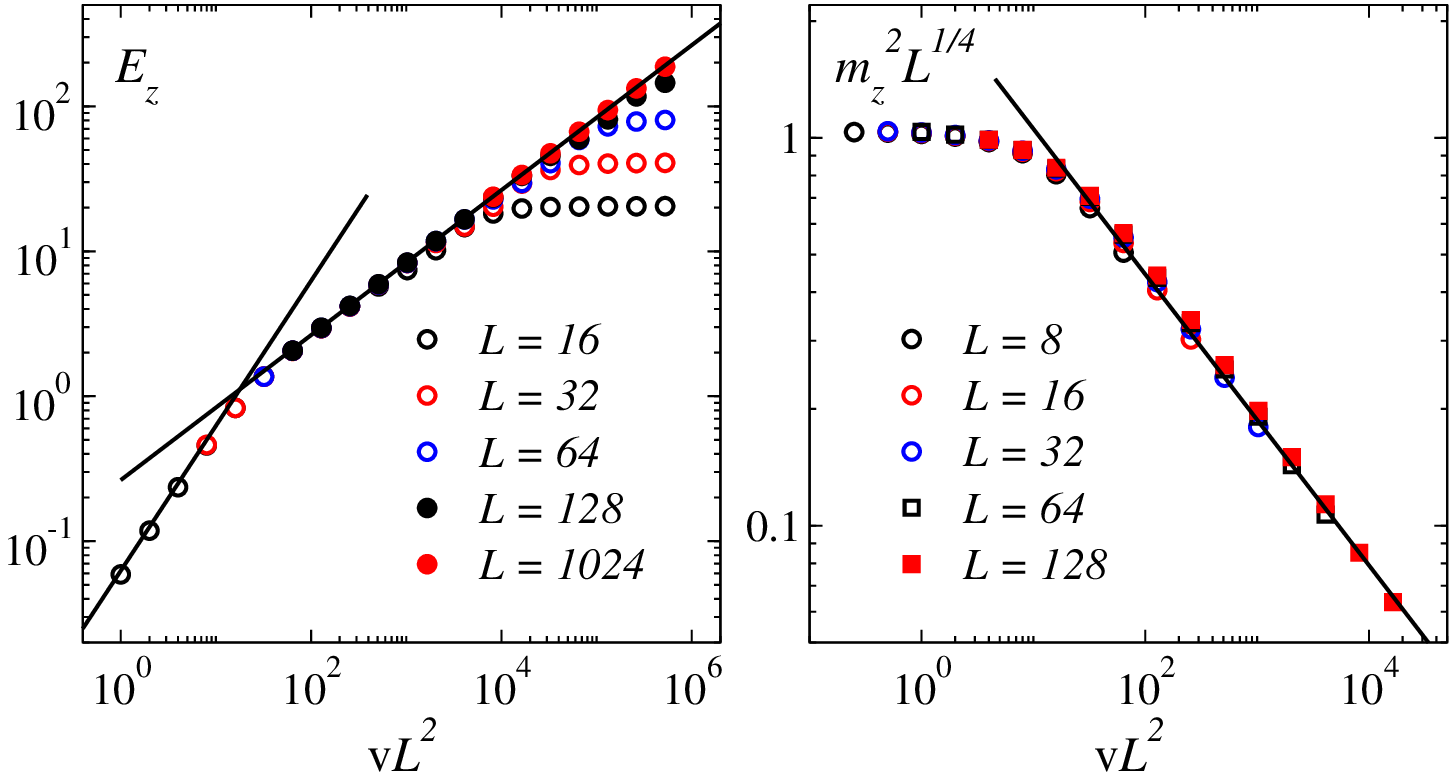}
\vskip-1mm
\caption{(Color online) Excess interaction energy $E_z$ (left) and squared magnetization (right) of Ising chains graphed according to our scaling predictions. 
The lines show the expected asymptotic forms; the slope is $1$ for $v L^2\ll1$ and $1/2$ for $v L^2\gg 1$ on the left and $-3/8$ for $v L^2\gg 1$ on the right.}
\label{plotM}
\end{center}
\vskip-4mm
\end{figure}

It is evident that the general scaling prediction (\ref{scaling}) indeed applies to this example. To illustrate further the finite-size scaling behavior predicted 
by Eq.~(\ref{scaling}),  in Fig.~\ref{plotM} (left panel) we plot the shift of the interaction energy with respect to the final ground state,
\be
E_z=-J\left[\sum_{\langle ij\rangle} \langle \sigma_z^i\sigma_z^j\rangle-\sum_{\langle ij\rangle} \langle 0|\sigma_z^i\sigma_z^j|0\rangle\right].
\label{ezdef}
\ee
versus $vL^2$. The data for different system sizes collapse, showing that one can use the proposed imaginary-time adiabatic approach to extract critical properties of a 
quantum phase transition.  To test our predictions further, we analyze the square of the longitudinal magnetization (the order parameter); 
\be
m_z^2=\frac{1}{L^2}\left \langle \left  (\sum_{j=1}^L \sigma_z^j \right )^2 \right \rangle, 
\ee
which has scaling dimension $\Delta_z=1/4$.\cite{sachdev} This together with Eq.~(\ref{scaling}) imply that 
\be
m_z^2 \approx L v^{5/8} f_{zJ}(L v^2)=L^{-1/4}\tilde f_{zJ}(vL^2). 
\ee
The large and small argument asymptotics of the scaling function $\tilde f_{zJ}$ are 
dictated by the equilibrium asymptotics in the diabatic limit, $\tilde f_{zJ}(x)\sim {\rm const}$ at $x\ll 1$, and by the requirement that $m_z^2\sim 1/L$ at 
$vL^2\gg 1$ when quenching from the disordered phase. If we quench from the ordered phase, $J(\tau_0)>1$, then $\tilde f_{zJ}(x)\sim x^{1/8}$ at $x\gg 1$ 
(so that $m_z^2\sim {\rm const}$). The finite-size scaling predictions and asymptotics are in excellent agreement with numerical data (Fig.~\ref{plotM}, left panel)
obtained using QMC simulations with the algorithm discussed next.

\section{Quantum Monte Carlo method}
\label{sec4}

A major advantage of the imaginary-time approach is that generalized QMC methods can be applied to evolve a state with the operator (\ref{utau}). Here we use an 
approach similar to the stochastic series expansion (SSE) method, discussed in the context of the transverse-field Ising model in Ref.~\onlinecite{Sandvik_03}. The method 
is generally applicable to all models for which standard equilibrium QMC simulations can be used, i.e., those for which there is no sign problem. Below we first
briefly review standard finite-temperature and ground-state QMC approaches. We then outline the general idea of the non-equilibrium QMC (NEQMC) method in imaginary 
time and apply it to the one- and two-dimensional transverse-field Ising models. 

\subsection{Standard QMC methods}

Standard QMC algorithms can be classified into finite-temperature methods, where the goal is to compute a quantum-mechanical
thermal average of the form
\begin{equation}
\langle A\rangle = \frac{1}{Z}{\bf Tr}\{ {\rm e}^{-\beta H}\},~~~~
Z = {\bf Tr}\{ {\rm e}^{-\beta H}\},
\end{equation}
and ground-state projector methods, where some operator $P(\beta)$ is applied to a ``trial state'' $|\Psi_0\rangle$, such that $|\Psi_\beta\rangle = P(\beta)|\Psi_0\rangle$ 
approaches the ground state when $\beta \to \infty$. Normally one is interested in expectation values,
\begin{equation}
\langle A\rangle = \frac{1}{Z}\langle \Psi_\beta|A|\Psi_\beta\rangle,~~~~ Z = \langle \Psi_\beta|\Psi_\beta\rangle,
\end{equation}
which approache the corresponding true ground state expectation values, $\langle A\rangle \to \langle 0| A|0\rangle$, when $\beta \to \infty$.
For the projector, one can use the imaginary-time evolution operator 
(\ref{utau}), $P(\beta)=U(\beta)={\rm e}^{-\beta \mathcal{H}}$, with a fixed (time-independent) Hamiltonian, or one can use a high power of the Hamiltonian, 
$P(\beta)=\mathcal{H}^m$, where $\beta \propto m/N$ gives the same rate of convergence (which is governed by the gap between the ground state and the first 
excited state in the symmetry sector of the trial state) for the two choices for a given system volume $N$. This follows from a Taylor expansion of the time 
evolution operator, which for large $\beta$ is dominated by powers of the order $n=\beta |E_0|$, where $E_0$ is the ground state energy (and $E_0 \propto N$) .

There are several ways to deal with the exponential. In the context of spins and bosons, the most frequently used methods are based on (i) the Suzuki-Trotter-decomposition, 
which leads to world-line methods,\cite{suzuki_77,hirsch_82} (ii) the continuous-time version of world-lines (e.g., the  worm algorithm\cite{prokofev_98}) and (iii) the Taylor 
expansion leading to the SSE method \cite{sandvik_91,sandvik_92,Sandvik_03} (see Ref.~\onlinecite{sandvik_10a} for a recent review of these approaches). The latter two methods 
are not affected by any approximations (beyond statistical sampling errors), while (i) has a discretization error. 

\subsection{Non-equilibrium QMC algorithm}

The NEQMC algorithm is similar to a ground-state projection, but instead of ${\rm e}^{-\beta \mathcal{H}}$ for a fixed Hamiltonian one uses the evolution operator (\ref{utau}) 
with a time dependent Hamiltonian. As in equilibrium QMC, one can treat the exponential operator in several different ways. Here we employ the series expansion.

Evolving from $\tau_0$ to $\tau$, Eq.~(\ref{utau}) is expanded in a power-series and applied to an initial state $|\Psi(0)\rangle$:
\beq
|\Psi(\tau)\rangle &=& \sum_{n=0}^\infty \int_{\tau_0}^\tau d\tau_n \int_{\tau_0}^{\tau_n} d\tau_{n-1} \cdots \int_{\tau_0}^{\tau_2} d\tau_1 \times \nonumber \\
&&~~~[\mathcal{-H}(\tau_n)] \cdots [\mathcal{-H}(\tau_1)]|\Psi(0)\rangle.
\label{qmc1}
\eeq
Writing $\mathcal{-H}$ in terms of individual site and bond operators, here denoted $H_{i}$, $i=1,\ldots,N_{\rm op}$, 
\be
\mathcal{-H} =\sum_{i=1}^{N_{\rm op}} H_i,
\ee
the operator product is written as a sum over all strings of these operators. Truncating at some maximum power $n=m$ (adapted to cause no detectable truncation
error, as in the SSE method\cite{sandvik_10a}) and introducing a trivial unit operator $H_{0}=1$, we can write Eq.~(\ref{qmc1}) as
\beq
|\Psi(\tau)\rangle &=& \sum_{H} \frac{(m-n)!}{(\tau - \tau_0)^{m-n}} 
\int_{\tau_0}^\tau d\tau_m \cdots \int_{\tau_0}^{\tau_3} d\tau_2  \int_{\tau_0}^{\tau_2} d\tau_1 \times \nonumber \\
&&~~~H_{{i_m}}(\tau_m) \cdots H_{{i_2}}(\tau_2) H_{{i_1}}(\tau_1)|\Psi(0)\rangle,
\label{qmc2}
\eeq
where $i_p \in \{0,\ldots, N_{\rm op}\}$, $\sum_{H}$ is the sum over all sequences $i_1,\ldots,i_m$, and $n$ is the number of indices $i_p \not =0$ in a given 
sequence. More generally, beyond the transverse-field Ising model, $i$ would refer to a lattice unit as well as a diagonal or off-diagonal part of the operator on 
this unit. The operators $H_i$ then have have the property that $H_i|\alpha\rangle = h_i(\alpha)|\alpha'\rangle$, where $|\alpha'\rangle$ is a basis state,
i.e., in the basis chosen to expand the states, there is no branching of the series of states obtained in the sequence of states resulting from the 
operators acting one-by-one in Eq.~(\ref{qmc2}).

As always in QMC simulations, we are in practicve restricted to systems for which the expansion is positive-definite, which is the same class for which sign problems can be avoided 
in equilibrium simulations. While the sign problem is a limitation of the QMC approach in general, the class of accessible models is still large and includes highly non-trivial 
and important systems. With the series expansion used in the NEQMC method here, avoiding the sign problem places constraints on the matrix elements $h_i(\alpha)$---the product 
of all matrix elements corresponding to a term in (\ref{qmc2}) has to be positive. 

\begin{figure}
\begin{center}
\includegraphics[width=8.4cm, clip]{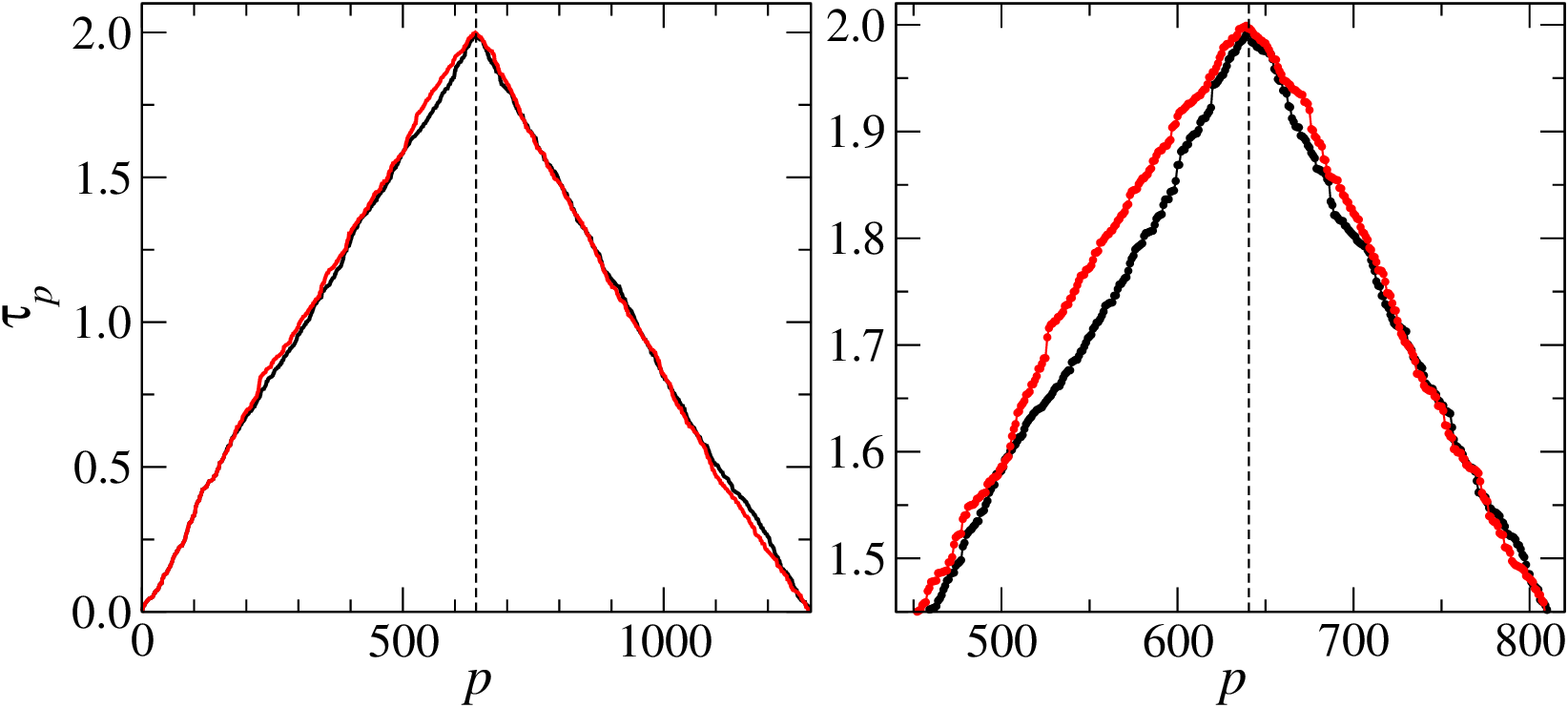}
\end{center}
\vskip-4mm
\caption{(Color online) Sampled imaginary-time sequences shown versus the propagation index $p$ after two successive Monte Carlo sweeps in a linear-quench simulation of 
an $8\times 8$ 2D transverse-field Ising model. At the initial time $\tau_0=0$ the Hamiltonian contains only the transverse field ($J=0,h=1$). The $J$-term is increased
linearly with time to the critical point $(J\approx 3.05,h=1)$ at the final time $\tau=2$. There is a total of $2m=1280$ operators, with $p=1,\ldots,640$ in a term of 
the projection of the ket state in (\ref{qmc1}) and $p=641,\ldots,1280$ in a corresponding bra term. The set of time points shown in red was obtained from those
shown in black by the method of updating several overlapping segments of a large number of times (here approximately 100) as discussed in the text. The right 
panel shows the behavior close to the center of the string (the final time) in greater detail.}
\label{tau}
\vskip-1mm
\end{figure}

Expectation values 
\be
\langle A \rangle_\tau =\frac{\langle \Psi(\tau)|A|\Psi(\tau)\rangle}{\langle \Psi(\tau)|\Psi(\tau)\rangle}
\ee
are computed by sampling the normalization
$\langle \Psi(\tau)|\Psi(\tau)\rangle$ written with (\ref{qmc2}). For the transverse-field Ising model, which we will apply the method to below, the method 
is very similar to the one developed in Ref.~\onlinecite{Sandvik_03} in the context of SSE QMC, the main difference being the change in the time boundaries; from periodic 
at finite temperature to those dictated by the initial state $|\Psi(0)\rangle$ of the time evolution. Changes in the operator sequence are made with the times $\tau_i$ 
fixed. The times are updated separately.

The operator sampling in the case of the transverse-field Ising model is particularly simple when the starting state is the equal superposition, 
\be
|\Psi(0)\rangle = \bigotimes_{i=1}^N |\uparrow_i + \downarrow_i\rangle,
\ee
which we use below, but other states can be used as well (in particular, for Heisenberg and other spin-isotropic
systems, amplitude-product states in the valence-bond basis \cite{liang_88} are very convenient, and a generalization of the loop updates used in the ground-state projector 
method of Ref.~\onlinecite{sandvik_10b} can be used). 

Since efficient operator and state cluster-updates  have been described in detail in the literature for various models in standard QMC simulations,\cite{Sandvik_03,sandvik_10b} 
only the time update (which is a generalization of a scheme previously discussed for equilibrium QMC in the interaction representation \cite{sandvik_97}) will be briefly 
outlined here. A whole segment of times, $\tau_i,\ldots,\tau_{i+n}$, can be simultaneously updated by generating $n+1$ numbers within the range $(\tau_{i-1},\tau_{i+n+1})$, 
then order these times according to a standard scheme scaling as $\log(n)$,\cite{numerical_07} and inserting the ordered set in place of the old segment of times. The 
Metropolis acceptance probability is easily obtained from (\ref{qmc2}), at a cost scaling as $n$. The number $n$ can be adjusted to give an acceptance probability close to 
$1/2$. Fig.~\ref{tau} shows an example of a time sequence and how it changes after a sweep of updates of partially overlapping segments covering the whole sequence of times.

\subsection{Results}

Using the NEQMC method we first confirmed that the exact results for the Ising chain are reproduced. Complete agreement was found to within
very small statistical errors. The results in the right panel of Fig.~\ref{plotM} are from the NEQMC simulations.

\begin{figure}
\begin{center}
\includegraphics[width=8.45cm, clip]{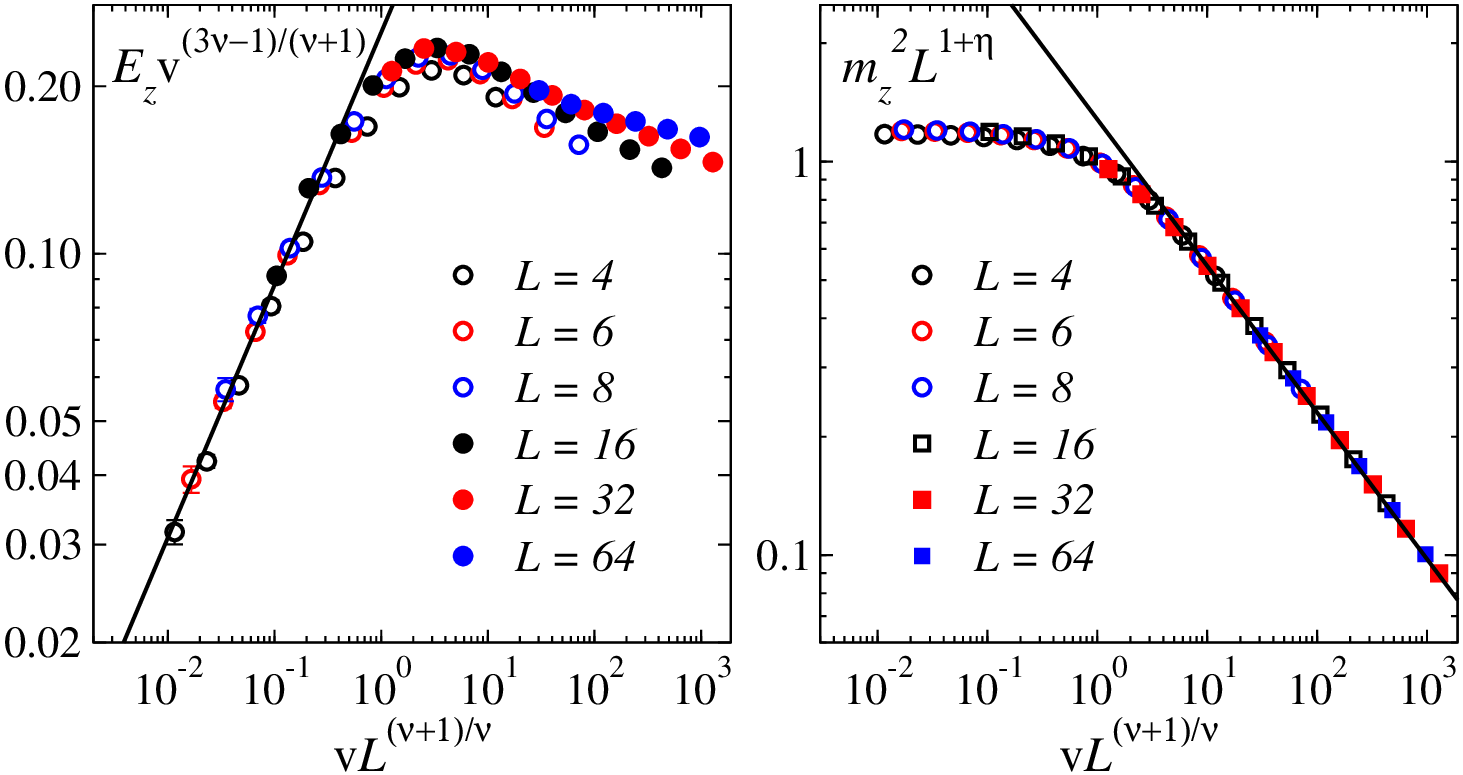}
\vskip-1mm
\caption{Same as Fig.~\ref{plotM} for the square-lattice model (quenching to its critical point  $J_c=0.32841$). QMC data are scaled according to the theoretical results, 
with the lines illustrating the predicted asymptotic slopes with exponents $z=1$, $\nu=0.6298$, and $\eta=0.0364$ (note that $\Delta_z=1+\eta$).\cite{Hasenbusch_99}}
\label{plotM2}
\end{center}
\vskip-4mm
\end{figure}

We next considered the same model on the 2D square lattice, i.e., the generalization of the 1D Hamiltonian (\ref{is1}). The critical coupling in this case is  $J_c=0.32841$ 
(based on exact diagonalization of a series of small lattices, which show behavior agreeing very well with predictions from low-energy field theory).\cite{hamer_00} 
In the left panel of Fig.~\ref{plotM2} we show the scaling of the excess Ising energy $E_z$, i.e., the 2D generalization of (\ref{ezdef}), 
for $L\times L$ lattices with $L$ up to $64$, using the known\cite{Hasenbusch_99} exponent $\nu=0.6298$ (obtained for the classical 3D Ising model, which 
should be in the same universality as the 2D quantum model studied here, which has dynamic exponent when $z=1$). We have divided $E_z$ by the leading powers of $L$ and $v$ 
predicted above and, hence, we should obtain a constant behavior for large $x$. This is not quite seen yet for these systems sizes, but the eventual convergence seems 
plausible. For smaller $x$ the data collapse very well and the asymptotic $x\to 0$ behavior is reproduced. In the right panel we show that also the squared magnetization 
scales according to our predictions, over five decades of the scaling argument $vL^{(\nu+1)/\nu}$. 

\section{Summary and discussion}
\label{sec5}

We have shown that detailed information on static and dynamic properties of a system can be obtained by propagating it in imaginary time. There are many 
similarities with real-time dynamics. In particular, we showed that one can use imaginary time to obtain universal exponents characterizing quantum critical 
points and to measure the fidelity susceptibility and components of the geometric tensor as response of physical observables to a linear quench. We obtained 
finite-size scaling expressions characterizing the response of various observables with the quench rate. In this way we extended the scaling theory of quantum 
phase transitions to non-equilibrium protocols. A clear advantage of the imaginary-time approach is that one can use powerful QMC simulations and 
circumvent complications related to real-time simulations. We have presented such a generic non-equilibrium QMC scheme and illustrated this approach for 
the transverse-field Ising model. Exact results (in one dimension) and QMC results (in one and two dimensions) show excellent agreement with the scaling 
predictions. The QMC method will be useful for studies of a wide range of non-trivial models on large lattices.

The ideas presented here apply also to quantum annealing, i.e., protocols where, in order to analyze the ground state of a complicated classical or quantum 
problem, one introduces an auxiliary coupling which makes the Hamiltonian simple and then slowly decreases this coupling to zero. This will allow one to address 
quantum annealing problems using QMC simulations.\cite{Liu_11}

\acknowledgments

We acknowledge useful discussions with V.~Gritsev. The work was supported by Grants NSF DMR-0907039 (AP and CDG), NSF DMR-0803510 and DMR-1104708 (AWS), 
AFOSR FA9550-10-1-0110 (AP), and the Sloan Foundation (AP).

\appendix

\section{Adiabatic perturbation theory}
\label{app1}

Let us discuss the leading non-adiabatic correction to the imaginary-time Schr\"odinger equation~(\ref{sch_eq}):
\be
\partial_\tau \psi(\tau)=-\mathcal H(\lambda(\tau))\psi.
\label{sch_eq1}
\ee
The natural way to address this question is to use adiabatic perturbation theory (APT), similar to that developed in Refs.~[\onlinecite{ortiz_2008, degrandi_2009}] 
in real time. We write the wave function in the instantaneous eigenbasis $\{ |n(\lambda)\rangle \}$ of $H(\lambda)$:
\be
\psi(\tau)=\sum_n a_n(\tau) |n(\lambda(\tau))\rangle.
\label{eq 1}
\ee
Substituting this expansion into Eq.~(\ref{sch_eq}) we find
\be
{d a_n\over d\tau}+\sum_m a_m(\tau) \langle n|\partial_\tau |m\rangle = -\mathcal E_n (\lambda) \, a_n(\tau),
\label{eq 2}
\ee
where $\mathcal E_n(\lambda)$ are the eigenenergies of $\cal H(\lambda)$ corresponding to the states $|n\rangle$. Making the 
transformation 
\begin{equation}
a_n(\tau)=\alpha_n(\tau)\exp\left[\int_\tau^0 \mathcal E_n(\tau')d\tau'\right],
\end{equation}
we can rewrite Eq.~(\ref{sch_eq}) as an integral equation [and note that $\alpha_n(0)=a_n(0)$]:
\beq
&&\alpha_n(\tau)=\alpha_n(0)+\sum_m \int^0_{\tau} d \tau'\, \langle n|\partial_{\tau'}|m\rangle \alpha_m(\tau')\nonumber\\
&&~~~~~~~~~~~\times\exp\left[-\int^0_{\tau'} d\tau''\, [\mathcal E_n(\tau'')-\mathcal E_m(\tau'')]\right].
\label{int_eq}
\eeq
In principle one should supply this equation with initial conditions at $\tau=\tau_0$ but, as we argued earlier, it is not necessary if $|\tau_0|$ is 
sufficiently large, since the sensitivity to the initial condition will be exponentially suppressed. Instead we impose the asymptotic condition 
$\alpha_n(\tau\to-\infty)\to \delta_{n0}$, implying that far in the past the system is effectively in the ground state.

Eq.~(\ref{int_eq}) is convenient for analysis with the APT. In particular, if the rate of change is very small, $\dot\lambda(\tau)\to 0$, 
then to leading order in $\dot\lambda$ the system remains in the ground state; $\alpha_m(\tau)\approx \delta_{m0}$ (except during the initial transient, 
which is unimportant at large $|\tau_0|$). In the next higher order the transition amplitudes to the states $n\neq 0$ are given by:
\be
\alpha_n(0)\approx -\int\limits^0_{-\infty} d\tau \, \langle n|\partial_{\tau}|0\rangle \exp\left[-\int^0_{\tau} d\tau'\, \Delta_{n0}(\tau')\right],
\label{eq_central1}
\ee
where $\Delta_{n0}(\tau)=\mathcal E_n(\tau)-\mathcal E_0(\tau)$.  The matrix element above for non-degenerate states can also be expressed as:
\be
\langle n|\partial_\tau|0\rangle =-\langle n|\partial_\tau \mathcal H(\tau)|0\rangle/ \Delta_{n0}(\tau).
\ee

\section{Adiabatic susceptibilities and non-equal time correlation functions}
\label{app2}

In this appendix we discuss the properties of the adiabatic susceptibilities [Eq.~(\ref{chi_g}) of the main text]:
\be
\chi_{\mu\lambda}^{(r+1)}={1\over L^d}\sum_{n\neq 0}{\langle 0|\partial_\lambda \mathcal H|n\rangle\langle n|\partial_\mu \mathcal H|0\rangle+\mu\leftrightarrow\lambda\over 2(\mathcal E_n-\mathcal E_0)^{r+1}}.
\ee
For linear quenches these quantities reduce to the symmetrized components of the geometric tensor\cite{venuti_07} up to a normalization factor. The representation of these 
susceptibilities through imaginary time correlation functions is a straightforward generalization of the result contained in Ref.~\onlinecite{venuti_07} (see also 
Ref.~\onlinecite{degrandi_10}):
\be
\chi_{\mu\lambda}^{(r+1)}\!\!={1\over 2 L^d}\int\limits_0^\infty d\tau {\tau^r\over r!} \langle 0| \partial_\mu \mathcal H_\tau\partial_\lambda \mathcal H_0+\partial_\lambda \mathcal H_\tau\partial_\mu \mathcal H_0|0\rangle_c,
\ee
where
\be
\partial_\lambda \mathcal H_\tau={\rm e}^{\tau \mathcal H} \partial_\lambda \mathcal H {\rm e}^{-\tau \mathcal H}.
\ee

Performing the Wick\rq{s} rotation $\tau\to it +\varepsilon$, where $\varepsilon$ is an infinitesimal positive number, we extend this result to real time:
\be
\chi_{\mu\lambda}^{(r+1)}\!\!={i^{r+1}\over 2 L^d}\int\limits_0^\infty dt {t^r\over r!} \langle 0| \partial_\mu \mathcal H_t\partial_\lambda \mathcal H_0+\partial_\lambda \mathcal H_t\partial_\mu \mathcal H_0|0\rangle_c,
\label{chi_1}
\ee
where 
\be
\partial_\lambda\mathcal  H_t={\rm e}^{it \mathcal H} \partial_\lambda \mathcal H {\rm e}^{-i t \mathcal H}
\ee
stands for the real-time Heisenberg operator. Thus we see that the adiabatic susceptibilities of order $r+1$ probe the $r$-th moment of the symmetric retarded correlation function of the operators $\partial_\lambda \mathcal H$ and $\partial_\mu \mathcal H$. Introducing the Fourier transform of this correlation function:
\be
G^{(r+1)}_{\mu\lambda}(\omega)=\int_0^\infty dt {\rm e}^{i\omega t} 
\langle 0| \partial_\mu \mathcal H_t\partial_\lambda \mathcal H_0+\partial_\lambda \mathcal H_t\partial_\mu \mathcal H_0|0\rangle_c,
\ee
we see that the susceptibility $\chi^{(r+1)}_{\mu\lambda}$ can be expressed through derivatives of the imaginary part of functions $G^{(r+1)}_{\mu\lambda}$ which define the structure 
factors:
\be
\chi_{\mu\lambda}^{(r+1)}=-{1\over 2r! L^d}{\partial^r\over \partial \omega^r} {\rm Im}\, G^{(r+1)}_{\mu\lambda}(\omega)\biggr|_{\omega=0}.
\ee

Finally let us mention the representation of these susceptibilities through the real part of non-equal time correlation functions. This can be achieved either by applying 
Kramers--Kronig relations to the equation above or directly from the definition:
\begin{widetext}
\be
\chi_{\mu\lambda}^{(r+1)}={1\over L^d}\int\limits_0^\infty d\omega \sum_{n\neq 0}{\langle 0|\partial_\lambda \mathcal H|n\rangle\langle n|\partial_\mu \mathcal H|0\rangle+
\mu\leftrightarrow\lambda\over 2\,\omega^{r+1}}\delta(\mathcal E_n-\mathcal E_0-\omega)={1\over 2 L^d}\int\limits_0^\infty {d\omega\over \omega^{r+1}}
{\rm Re}\,G^{>(r+1)}_{\mu\lambda}(\omega),
\ee
\end{widetext}
where
\be
G^{>(r+1)}_{\mu\lambda}(\omega)=\int_{-\infty}^\infty dt \mathrm e^{i\omega t}
\langle 0|\partial_\lambda \mathcal H_t \partial_\mu \mathcal H_0+\lambda\leftrightarrow\mu|0\rangle.
\ee

\bibliography{adiabatic_it}

\end{document}